\documentclass[aip,amsmath,amssymb,reprint]{revtex4-2}
\usepackage{graphicx}
\usepackage{color}
\usepackage{bm}
\usepackage[hidelinks]{hyperref}

\newcommand*{\diff}{\mathop{}\!\mathrm{d}}

\renewcommand*{\vec}[1]{\mathbf{#1}}

\renewcommand*{\tensor}[1]{\boldsymbol{#1}}

\graphicspath{{./figures/}}

\begin{document}

\title{Inhomogeneous steady shear dynamics of a three-body colloidal gel former}

\author{Florian Sammüller}
\author{Daniel de las Heras}
\author{Matthias Schmidt}
\email{Matthias.Schmidt@uni-bayreuth.de}
\affiliation{Theoretische Physik II, Physikalisches Institut, Universität Bayreuth, D-95447 Bayreuth, Germany}

\date{\today}

\begin{abstract}
  We investigate the stationary flow of a colloidal gel under an inhomogeneous external shear force using adaptive Brownian dynamics simulations.
  The interparticle forces are derived from the Stillinger-Weber potential, where the three-body term is tuned to enable network formation and gelation in equilibrium.
  When subjected to the shear force field, the system develops remarkable modulations in the one-body density profile.
  Depending on the shear magnitude, particles accumulate either in quiescent regions or in the vicinity of maximum net flow, and we deduce this strong non-equilibrium response to be characteristic of the gel state.
  Studying the components of the internal force parallel and perpendicular to the flow direction reveals that the emerging flow and structure of the stationary state are driven by significant viscous and structural superadiabatic forces.
  Thereby, the magnitude and nature of the observed non-equilibrium phenomena differs from the corresponding behavior of simple fluids.
  We demonstrate that a simple power functional theory reproduces accurately the viscous force profile, giving a rationale of the complex dynamical behavior of the system.
\end{abstract}

\maketitle

\section{Introduction}

Gelation in soft matter is a complex and important phenomenon that has many practical applications ranging from the use in household materials to advanced technological processes \cite{Chen2011,Wang2012,Sacanna2013,Demortiere2014,Sherman2021,Lattuada2022}.
A common property of gels is their ability to sustain weak external stresses due to the formation of persistent long-range network structures.
From a microscopic point of view, it is the nontrivial correlation of particles arising from their internal interactions that give gels their characteristic mechanical response \cite{Dinsmore2002,Stokes2008,Laurati2009}.

However, the route to the generation of the network topology can be diverse \cite{Zaccarelli2007}.
One common path to the gelation of colloids involves the crossing of a liquid-gas spinodal, e.g.\ by a sudden quench in temperature, and the subsequent dynamical arrest of heterogeneous dense regions.
This arrested spinodal decomposition is non-equilibrium in nature and it bears similarity to the glass transition, although it is driven by interparticle attraction rather than by repulsion \cite{Cates2004,Royall2008,Royall2018}.
On the other hand, an equilibrium route to gel formation lies open by careful choice of the interparticle interactions in order to prevent macroscopic liquid-gas phase separation and to favor instead the local arrangement of particles into interconnected clusters or chains.
In this spirit, a multitude of interaction potentials have been investigated, which incorporate, for example, limitation of particle connectivity \cite{Zaccarelli2005,Zaccarelli2006,Lindquist2016,Howard2021}, competing short-range attraction and long-range repulsion \cite{Groenewold2004,Suarez2009}, and anisotropy \cite{Blaak2007,Miller2009,Rovigatti2011} via, e.g., ``patchy'' interaction sites \cite{Bianchi2011,DelGado2010,Sciortino2011}.
The liquid-gas spinodal can sometimes be pushed to very low temperatures and densities, which enables large parts of the phase diagram to be governed by the percolation into dilute networks, as in so called ``empty liquids'' \cite{Bianchi2006,Ruzicka2011,delasHeras2011}.

A further class of particle models for colloidal gels that the present work focuses on is based on the inclusion of three-body interactions to the interparticle interaction potential, which consists otherwise only of isotropic pair-interactions \cite{Saw2009,Saw2011,Colombo2013,Colombo2014,Colombo2014a,Hatami-Marbini2020,Hatami-Marbini2020a,Bantawa2021}.
It has been shown that an appropriate choice of the three-body term reproduces the distinctive network topology \cite{Bouzid2018} as well as the characteristic non-linear response to homogeneous shear, including strain hardening and yielding \cite{Colombo2014a}.
Especially under external load, the dynamics of such a gel can be intricate, e.g.\ exhibiting cooperative restructuring of particle bonds \cite{Colombo2013} and shear banding \cite{Colombo2014a}.

While many studies have considered the response of gels to a linear shear profile up to their breaking point, not much is known about their viscous flow behavior in \emph{inhomogeneous} external force fields.
However, it can be expected that the intrinsic features of a gel former, such as the tendency of particles to percolate, have substantial ramifications in such out-of-equilibrium scenarios.
Specifically, one is tempted to assume that some of the genuine non-equilibrium effects \cite{Stuhlmuller2018,delasHeras2020} already reported for simple fluids (such as shear migration) might even be amplified by additional three-body interactions.

In this work, we show that gels modeled via a modified Stillinger-Weber \cite{Stillinger1985} potential with a preferred three-body angle of $180^\circ$ as proposed by Saw, Ellegaard, Kob, and Sastry (SEKS) \cite{Saw2009,Saw2011} are indeed highly susceptible to these non-equilibrium effects when sheared by a sinusoidal external force profile.
For this, we numerically investigate the behavior of the SEKS model with adaptive Brownian dynamics \cite{Sammuller2021} (adaptive BD), which is a stable and efficient method for the simulation of many-body systems governed by the overdamped Langevin equations of motion.
We find that the properties of the emerging stationary state vary strongly with temperature and with the amplitude of the external force profile.
Different behavior occurs in the shape of both the density and the internal force profiles as compared to simple fluids.
In particular, we show that the superadiabatic (i.e.\ genuine out-of-equilibrium) contribution to the internal force is substantial in magnitude and that it is responsible for the structural and viscous behavior of the stationary shear flow.
This is discussed from a microscopic point of view as well as in a coarse-grained fashion, where we use power functional theory \cite{Schmidt2013,Schmidt2022} (PFT) to develop a quantitative model for the superadiabatic viscous force.
Besides representing a generic situation, the sinusoidal shear flow profile could be seen as a toy model for a mesoscopic convection roll.
Convection typically occurs in sedimentation as upward streams alternate with downward streams \cite{Royall2007}.

This work is structured as follows.
In Sec.~\ref{sec:particle_model}, the modified Stillinger-Weber potential as well as details for its efficient computation are given.
The adaptive BD method and its advantages for our non-equilibrium simulations are laid out in Sec.~\ref{sec:adaptive_brownian_dynamics}.
In Sec.~\ref{sec:simulation_protocol}, the protocol for the simulation of the stationary flow state is described.
In Secs.~\ref{sec:T_sweep} and \ref{sec:K_sweep}, we show one-body profiles of the density as well as the parallel and perpendicular components of the internal force for a range of simulation parameters, and discuss their behavior and interplay.
An analogous interpretation on the level of internal stresses is given in Appendix \ref{appendix:stress}.
In Appendix \ref{appendix:theta0}, we showcase results for different values of the three-body angle of the Stillinger-Weber potential, and in Appendix \ref{appendix:LJ} the unusual nonequilibrium response of the gel is contrasted with numerical results for the simple Lennard-Jones fluid.
In Sec.~\ref{sec:PFT}, the description of superadiabatic forces with PFT is illustrated and the results are compared with those from simulation.
We conclude in Sec.~\ref{sec:conclusion_and_outlook} and give an outlook to the investigation of further dynamical phenomena observed in our simulations and to a more extensive analysis with PFT.

\section{Simulation method}

\subsection{Particle model}
\label{sec:particle_model}

The Stillinger-Weber potential \cite{Stillinger1985} has originally been used for the simulation of solid and liquid silicon, and it has since been optimized and adapted to other particle types \cite{Barnard2002,Bhat2007}.
The interparticle interactions consist of a two-body potential $u_2(r)$ that models both isotropic attraction and repulsion depending on the distance $r$ between two particles, as well as a three-body contribution $u_3(r, r', \Theta)$.
This three-body term imposes an energetically favorable angle $\Theta$ for three particles where a central particle is separated by the pairwise distances $r$ and $r'$ to two other particles.
The directionality of internal interactions is therefore only realized via $u_3$.
Crucially, there is no need to explicitly incorporate orientational degrees of freedom, which is an advantage both in simulations as well as in a theoretical treatment.

In total, the internal energy potential possesses the form
\begin{equation}
  \label{eq:SW}
  U(\vec{r}^N) = \sum^N_i \sum^N_{j > i} u_2(r_{ij}) + \sum^N_i \sum^N_{j \neq i} \sum^N_{k > j} u_3(r_{ij}, r_{ik}, \Theta_{ijk})\\
\end{equation}
with
\begin{gather}
  \label{eq:u2}
  u_2(r) = A \epsilon \left[ B \left( \frac{\sigma}{r} \right)^p - \left( \frac{\sigma}{r} \right)^q \right] \exp\left(\frac{\sigma}{r - a \sigma}\right),\\
  \label{eq:u3}
  \begin{split}
    u_3(r, r', \Theta) &= \lambda \epsilon \left[ \cos\Theta - \cos\Theta_0 \right]^2\\
                       &\qquad \times \exp\left(\frac{\gamma \sigma}{r - a \sigma}\right) \exp\left(\frac{\gamma \sigma}{r' - a \sigma}\right),
  \end{split}
\end{gather}
for a certain particle configuration $\vec{r}^N = \{\vec{r}^{(i)}, \dots, \vec{r}^{(N)}\}$ of the many-body system with $N$ particles.

The parameters $p$, $q$, $A$, $B$, $a$, $\gamma$, $\lambda$ and $\Theta_0$ can be tuned to alter the shape of the potential.
A choice for these quantities, which is used in the present work and varies in some aspects from the one used originally by \citeauthor{Stillinger1985} \cite{Stillinger1985}, is given in Table \ref{tab:SW_parameters}.
In particular, following previous works of \citeauthor{Saw2009} \cite{Saw2009,Saw2011}, we tune $\Theta_0$ to obtain a gel former, which is described in more detail below.
The formulation in eqs.~\eqref{eq:u2} and \eqref{eq:u3} refrains from using absolute units and only involves intrinsic energy ($\epsilon$) and length ($\sigma$) scales.
In an overdamped system with friction coefficient $\zeta$, all physical quantities can therefore be expressed in a reduced form.

We note that the parameter $a$ sets the cutoff distance since both $u_2(r)$ and $u_3(r, r', \Theta)$ as well as their gradients vanish smoothly for $r \rightarrow a \sigma$ and $r' \rightarrow a \sigma$.
The potential is therefore inherently short-ranged (cf.\ the small value of $a$ in Table \ref{tab:SW_parameters}).
This is a favorable propery for the treatment in computer simulations since it enables the use of neighbor-tracking algorithms to avoid superfluous evaluations for particles beyond the cutoff distance, which substantially reduces the computational cost in large systems.

The parameter $\Theta_0$ in the three-body term $u_3(r, r', \Theta)$ sets the preferred angle of a certain particle triplet (note that $u_3$ vanishes for $\Theta = \Theta_0$ and that it is otherwise strictly positive for particles within the cutoff distance).
Most commonly, as discussed below, tetrahedral configurations are desired, for which one chooses $\cos\Theta_0 = - 1/3$.
The strength of the three-body interaction term is adjusted via $\lambda$, which is often referred to as the tetrahedrality \cite{Molinero2009} for the above choice of $\Theta_0$.

A further computational optimization is employed, which makes use of the concrete structure of $u_3(r, r', \Theta)$ as given in eq.~\eqref{eq:u3}.
Via a rewriting of the three-body sum and the introduction of accumulation variables, an evaluation of the total energy and of all particle forces is possible by only iterating twice over all interacting particle \emph{pairs}.
In contrast, a naive implementation would require an iteration over particle triplets.
For details of this exact reformulation, which leads to a substantial speedup in our simulations \footnote{The simulation code can be found at \url{https://gitlab.uni-bayreuth.de/bt306964/mbd}.}, consult Ref.~\onlinecite{Saw2011}.

In summary, the versatility and computational efficacy of the Stillinger-Weber potential make it applicable to a wide range of problems.
An important example, which conveys its use as an \emph{effective} interaction potential for more complex particle types, is the monatomic water model of \citeauthor{Molinero2009} \cite{Molinero2009}.
It has been shown by these authors that thermodynamic and structural properties of water (e.g.\ for the study of interfacial phenomena \cite{Coe2022}) can be captured accurately by this model via an appropriate choice of the absolute values of $\epsilon$ and $\sigma$ as well as the tetrahedrality $\lambda$.
By comparison with the melting temperature of water, they determined an optimal value of $\lambda = 23.15$, which lies between the respective tetrahedralities of silicon and carbon and which is adopted in our simulations.

While eq.~\eqref{eq:u3} has initially been conceptualized as a model for tetrahedrally coordinated particles, it is entirely conceivable to alter the preferred three-body angle $\Theta_0$.
A variation of $\Theta_0$ has significant consequences for the spatial correlations of the fluid, since the formation of droplets might become energetically unfavorable and the self-assembly into interconnected chains that form open networks is enforced.
Therefore, the careful choice of the values of $\Theta_0$ and $\lambda$ is a means to reduce the effective valency and to suppress the liquid-gas phase transition, making the Stillinger-Weber potential \eqref{eq:SW} a suitable model for colloidal gels.
In the following, we set $\Theta_0 = 180^\circ$, although other values of $\Theta_0$ have been shown to support gelation as well, e.g.\ as reported in Refs.\ \onlinecite{Saw2009,Saw2011}, where a detailed investigation of the phase diagram and percolation behavior was carried out for various choices of $\lambda$ and $\Theta_0$.
(In Appendix \ref{appendix:theta0}, illustrative results are presented for lower values of $\Theta_0$, which shows that its precise value has little impact on the sheared steady state as long as network formation can occur.)
It is worth noting that gelation has also been investigated for other choices of two- and three-body interaction terms $u_2$ and $u_3$ apart from those given in eqs.~\eqref{eq:u2} and \eqref{eq:u3}, see e.g.\ Refs.~\onlinecite{Colombo2013,Colombo2014,Colombo2014a,Hatami-Marbini2020,Hatami-Marbini2020a,Bantawa2021}.
For instance, to yield stronger angular rigidity, the cosine difference in eq.~\eqref{eq:u3} can been exponentiated \cite{Colombo2013,Colombo2014,Colombo2014a}.

\begin{table}[htb]
  \caption{In eqs.~\eqref{eq:u2} and \eqref{eq:u3}, we adopt the parameters $p$, $q$, $A$, $B$, $a$ and $\gamma$ of the original Stillinger-Weber \cite{Stillinger1985} potential and choose the three-body strength $\lambda$ as determined in Ref.~\onlinecite{Molinero2009}.
  In accordance with \citeauthor{Saw2009} \cite{Saw2009,Saw2011}, a preferred three-body angle of $\Theta_0 = 180^\circ$ then leads to the percolation of inter-connected chains, enabling colloidal gelation in equilibrium.}
  \label{tab:SW_parameters}
  \begin{tabular}{|c|c|c|c|c||c|c|c|}
    \hline
    $p$ & $q$ & $A$ & $B$ & $a$ & $\gamma$ & $\lambda$ & $\Theta_0$ \\
    \hline
    4 & 0 & 7.04955627 & 0.6022245584 & 1.8 & 1.2 & 23.15 & $180^\circ$ \\
    \hline
  \end{tabular}
\end{table}

\subsection{Adaptive Brownian dynamics}
\label{sec:adaptive_brownian_dynamics}

An important property of gels is their mechanical response to externally imposed strain.
As particle bonds within the network are capable of sustaining substantial forces and torques without breaking, a gel exhibits elastic behavior before stiffening \cite{Pouzot2006} as well as yielding at intermediate and large shear strain due to bending and breaking of bonds respectively \cite{Colombo2014a}.
Numerically, these results can be obtained, e.g., by performing a linear deformation of the simulation box and measuring the stress tensor.
When using nonequilibrium molecular dynamics, adequate thermostatting is required \cite{Evans1984,Todd2017}, which is not straightforward if spatially inhomogeneous deformations are considered.
This is even more problematic if cause and effect are reversed, and an external force profile is applied which generates a macroscopic net flow that is hence not known a priori.
We circumvent these issues by considering overdamped dynamics, where thermostatting is intrinsic.

Furthermore, an advanced numerical integration scheme known as adaptive BD \cite{Sammuller2021} is applied, which improves upon conventional BD simulations as described in the following.
We consider the overdamped Langevin equations
\begin{equation}
  \label{eq:overdamped_Langevin}
  \dot{\vec{r}}^{(i)}(t) = \frac{1}{\zeta} \vec{f}^{(i)}(\vec{r}^N(t)) + \sqrt{\frac{2 k_B T}{\zeta}} \vec{R}^{(i)}(t),
\end{equation}
$i = 1, \dots, N$, as the relevant equations of motion to obtain particle trajectories $\vec{r}^N(t)$ in our system consisting of $N$ identical particles.
Here, $\vec{f}^{(i)}(\vec{r}^N(t))$ is the total force acting on particle $i$, which can be split into external and internal contributions, $\vec{f}^{(i)}_\mathrm{ext}(\vec{r}^{(i)}(t))$ and $\vec{f}^{(i)}_\mathrm{int}(\vec{r}^N(t)) = - \nabla_i U(\vec{r}^N(t))$, respectively.
The friction coefficient $\zeta$ is the same for each (identical) particle and the over dot denotes a time derivative.
The vectors $\vec{R}^{(i)}(t)$, $i = 1, \dots, N$, are Gaussian distributed and must therefore satisfy $\langle \vec{R}^{(i)}(t) \rangle = 0$ and $\langle \vec{R}^{(i)}(t) \vec{R}^{(j)}(t') \rangle = \tensor{I} \delta_{ij} \delta(t - t')$.
Here, the angular brackets denote an average over realizations of the random process, $\tensor{I}$ is the $3 \times 3$-unit-matrix, $\delta_{ij}$ is the Kronecker delta and $\delta(\cdot)$ is the Dirac delta function.
Thermostatting irrespective of applied external forces and any (possibly inhomogeneous) net flow is inherent in overdamped Brownian dynamics, as the temperature-dependent prefactor of the Gaussian random vectors in eq.~\eqref{eq:overdamped_Langevin} determines the average magnitude of the random displacements.

Because eq.~\eqref{eq:overdamped_Langevin} is a set of coupled stochastic differential equations, its numerical treatment requires particular care.
Specifically, the use of the Euler-Maruyama method \cite{Kloeden1999}, which is usually employed in conventional BD simulations, has serious drawbacks regarding both its stability and accuracy.
This is primarily due to using a constant timestep interval $\Delta t$, which may lead to faulty particle displacements and erroneous force evaluations when particle collisions are not resolved with the required precision (i.e.\ with a small enough $\Delta t$).

Within adaptive BD \cite{Sammuller2021}, the automatic choice of an appropriate timestep length $\Delta t_k$ is ensured in each iteration $k \rightarrow k + 1$ by the evaluation of an embedded Heun-Euler integrator
\begin{align}
  \label{eq:Euler}
  \bar{\vec{r}}^{(i)}_{k + 1} &= \vec{r}^{(i)}_k + \frac{1}{\zeta} \vec{f}^{(i)}(\vec{r}^N_k) \Delta t_k + \sqrt{\frac{2 k_B T}{\zeta}} \vec{R}^{(i)}_k,\\
  \label{eq:Heun}
  \begin{split}
    \vec{r}^{(i)}_{k + 1} &= \vec{r}^{(i)}_k + \frac{1}{2 \zeta} \left( \vec{f}^{(i)}(\vec{r}^N_k) + \vec{f}^{(i)}(\bar{\vec{r}}^N_{k + 1}) \right) \Delta t_k\\
                          &\qquad + \sqrt{\frac{2 k_B T}{\zeta}} \vec{R}^{(i)}_k,
  \end{split}
\end{align}
which yields two estimates $\bar{\vec{r}}^N_{k + 1}$ and $\vec{r}^N_{k + 1}$ for the new particle positions at time $t_k + \Delta t_k$.
If large discrepancies of $\bar{\vec{r}}^N_{k + 1}$ and $\vec{r}^N_{k + 1}$ are detected, the timestep $\Delta t_k$ is reduced and the step $k \rightarrow k + 1$ is retried.
In such a case of a rejected trial step, one must carefully choose appropriate discrete random increments $\vec{R}^{(i)}_k$ to retain the Gaussian nature of the target random process $\vec{R}^{(i)}(t)$.
For this, adaptive BD utilizes Rejection Sampling with Memory (RSwM) \cite{Rackauckas2017}, which is an efficient algorithm to counteract the rejection of previously drawn random increments.
RSwM hence guarantees the correct generation of a specified random process.
With the numerical treatment of eq.~\eqref{eq:overdamped_Langevin} via adaptive BD, stable and accurate long-time simulations of overdamped many-body systems are possible in equilibrium but also under extreme non-equilibrium conditions, such as when driving the system with a large external force $\vec{f}_\mathrm{ext}(\vec{r})$.
A detailed description of adaptive BD is given in Ref.~\onlinecite{Sammuller2021}.

Hydrodynamic interactions that are mediated by the implicit solvent are neglected in the equations of motion \eqref{eq:overdamped_Langevin}.
From a computational standpoint, the performance of the simulations turns out to be crucial to obtain accurate results for the quantities of interest, as described in the next section.
The numerical treatment of hydrodynamic interactions, which requires both the evaluation of long-ranged forces as well as the generation of appropriately correlated random displacements (e.g.\ via a Cholesky decomposition of the diffusion matrix \cite{Ermak1978}), would have a significant impact on the required computational effort.
Additionally, within adaptive BD, the Heun-Euler pair \eqref{eq:Euler} and \eqref{eq:Heun} is conceived to handle only additive noise in the underlying stochastic differential equation.
Instead of the Heun method \eqref{eq:Heun}, one would have to resort to an integration scheme with a sufficient strong order of convergence for general noise terms \cite{Kloeden1999}.
Moreover, from a physical point of view, we argue that the omission of hydrodynamic interactions simplifies the analysis of the results in Sec.~\ref{sec:results}, as all observations are ensured to stem solely from the properties of the Stillinger-Weber particle model.
In particular, we show that its ability to form networks is the crucial mechanism that causes the reported out-of-equilibrium response.
As hydrodynamic interactions tend to support anisotropic coagulation and transient network states \cite{Tanaka2000,Kovalchuk2012,Royall2015}, we expect no significant qualitative change in the reported observations.

\subsection{Simulation protocol}
\label{sec:simulation_protocol}

\begin{figure*}[tb]
  \centering
  \includegraphics[width=\textwidth]{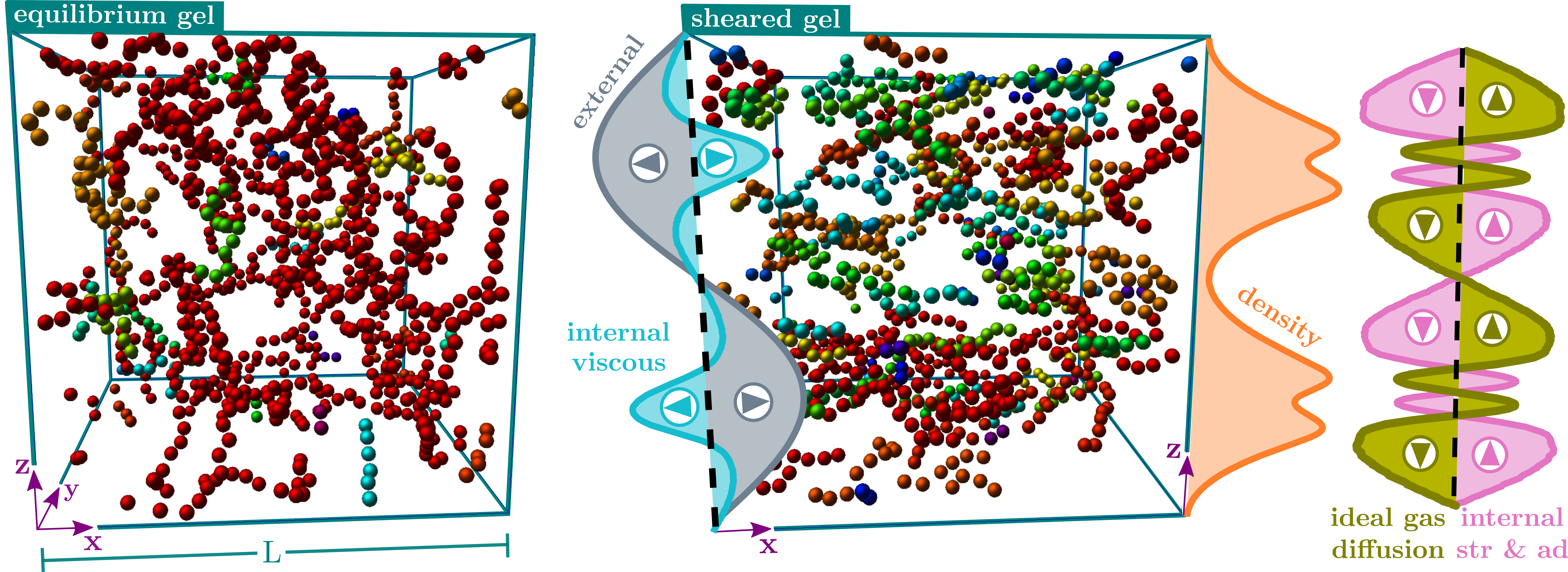}
  \caption{
    Characteristic snapshots of an equilibrium gel (left) and a sheared gel in steady-state (right)\footnote{An animation of the sheared gel can be found at \url{http://www.staff.uni-bayreuth.de/~bt306964/gelation/sheared_gel_T0.1_K5.mp4}.}.
    The particles are colored according to the cluster to which they belong.
    The simulation box is a cube of side length $L$ and the temperature is set to $k_B T = 0.1 \epsilon$.
    For the sheared state, an external force amplitude of $K = 5 \epsilon / \sigma$ is chosen.
    The forces acting on the sheared gel are schematically represented:
    A sinusoidal external force pointing along the $x$-direction drives the flow of particles, forming two flow channels with the velocity profile (gray) closely following the external force profile \eqref{eq:fext}.
    In steady state, a superadiabatic viscous internal force (cyan) emerges that locally either opposes or supports the flow.
    A strong density modulation (orange) develops along the $z$-direction.
    The ideal gas diffusive force (olive) that tends to homogenize the density profile is balanced by an internal force (pink) along the $z$-direction, which incorporates both adiabatic and superadiabatic structural components.
  }
  \label{fig:system_sketch}
\end{figure*}

In this work, we model an inhomogeneously sheared system by imposing an external force profile $\vec{f}_\mathrm{ext}(z)$ that is parallel to the $x$-axis and modulated in the $z$-direction.
The $x$-component of the force field is sinusoidal with amplitude $K$ such that it complies with the periodic boundary conditions of the cubic simulation box with side length $L$, i.e.\
\begin{equation}
  \label{eq:fext}
  f_{\mathrm{ext}, x}(z) = K \sin\left(2 \pi \frac{z}{L}\right).
\end{equation}

With this choice, Lees-Edwards boundary conditions \cite{Lees1972} as used in simulations with linear shear profiles \cite{Jahreis2020} are not required, because $\vec{f}_\mathrm{ext}(\vec{r})$ as well as its derivatives are continuous at the periodic boundaries.
The application of the time-independent but spatially inhomogeneous shear force \eqref{eq:fext} is not to be confused with time-dependent oscillatory shear \cite{Moghimi2017}.
The considered external force constitutes the lowest-order Fourier mode within the simulation box, and it can hence be taken as a generic model for experimentally relevant scenarios, as occur e.g.\ in convection \cite{Royall2007} or when inducing inhomogeneous forces with a laser tweezer.

The simulation procedure is as follows.
We set $L = 30 \sigma$ and initialize $N = 1000$ particles on a regular lattice, which yields a mean number density of $\rho_b \approx 0.037 \sigma^{-3}$.
This configuration is randomized for a short time ($10^4$ steps) at a high temperature of $k_B T = 10 \epsilon$ using the adaptive BD method, before instantaneously reducing the temperature to the desired value and imposing the sinusoidal external force profile \eqref{eq:fext}.
At this point, no particle bonds have formed yet and a flow in the $x$-direction sets in immediately.
From here, the actual production run begins and the respective observables are sampled, which is described in more detail below.
Due to the nature of the external force profile \eqref{eq:fext}, the system retains translational invariance in the $x$-$y$-plane and forms a flow channel in the upper and lower half of the simulation box respectively.
Since a transient from the randomized particle distribution into this stationary flow occurs initially, we partition the sampling of the production run into consecutive sections of $10^6$ steps.
Thus, the sections where a stationary flow has not been reached yet can be discarded, and the remaining ones are averaged over.
During individual runs, asymmetric channel populations that persist for a long time are observed.
Rather than performing longer simulation runs to yield better time-averages, we average over approximately $50$ distinct realizations of a simulation until symmetric profiles are obtained.
The typical simulation time of the stationary flow in each individual run is then in the order of $10^5 \tau$ with the Brownian timescale $\tau = \sigma^2 \zeta / \epsilon$.

Using this protocol, a range of external modulation amplitudes $K$ and temperatures $T$ is investigated.
For each set of parameters, we obtain the density profile $\rho(\vec{r})$ as well as the force density profile $\vec{F}(\vec{r})$ from sampling of the density operator
\begin{equation}
  \label{eq:rho_op}
  \hat{\rho}(\vec{r}) = \sum_i \delta(\vec{r} - \vec{r}^{(i)})
\end{equation}
and force density operator
\begin{equation}
  \label{eq:F_op}
  \hat{\vec{F}}(\vec{r}) = \sum_i \vec{f}^{(i)} \delta(\vec{r} - \vec{r}^{(i)}),
\end{equation}
respectively.
Thus, $\rho(\vec{r}) = \langle \hat{\rho}(\vec{r}) \rangle$ and $\vec{F}(\vec{r}) = \langle \hat{\vec{F}}(\vec{r}) \rangle$, where angular brackets denote an average over configurations of the stationary flow state obtained according to the above simulation procedure.
Specifically, for the force density profile, we focus on its internal contribution
\begin{equation}
  \label{eq:Fint}
  \vec{F}_\mathrm{int}(\vec{r}) = \left\langle \sum_i \vec{f}^{(i)}_\mathrm{int} \delta(\vec{r} - \vec{r}^{(i)}) \right\rangle
\end{equation}
to better reveal how the stationary state is stabilized by the internal interaction \eqref{eq:SW}.
The internal force density profiles are then normalized by the density to acquire the internal force profile
\begin{equation}
  \label{eq:fint}
  \vec{f}_\mathrm{int}(\vec{r}) = \frac{\vec{F}_\mathrm{int}(\vec{r})}{\rho(\vec{r})}.
\end{equation}

A sufficiently large number of samples is necessary to yield accurate results for the internal force profile, as its convergence is slower than that of the density profile \footnote{For one steady state, the profiles were obtained within 1000 CPU hours.}.
To investigate possible finite-size effects, which might occur in gels specifically due to their long-range effective correlations, we have conducted additional simulations where the side length $L$ of the box has been doubled while extending the external potential \eqref{eq:fext} in the $z$-direction by an additional shear period.
No significant impact was found on the behavior of the sheared system compared to the results shown in the next section for the original choice of $L$.

A sketch of the system and of the flow velocity profile resulting from the applied shear force \eqref{eq:fext} is depicted in fig.~\ref{fig:system_sketch}, where we also show characteristic snapshots of the quiescent and of the sheared gel.
Additionally, the spatial variations of the one-body profiles are illustrated and we indicate locally by arrows the directions of the one-body force contributions.
Actual simulation results are presented and analyzed in the following, and we highlight the labels of the one-body profiles in subsequent figures according to the colors used in fig.\ \ref{fig:system_sketch}.

\section{Results}
\label{sec:results}

\subsection{Variation of temperature}
\label{sec:T_sweep}

\begin{figure}[htb]
  \centering
  \includegraphics{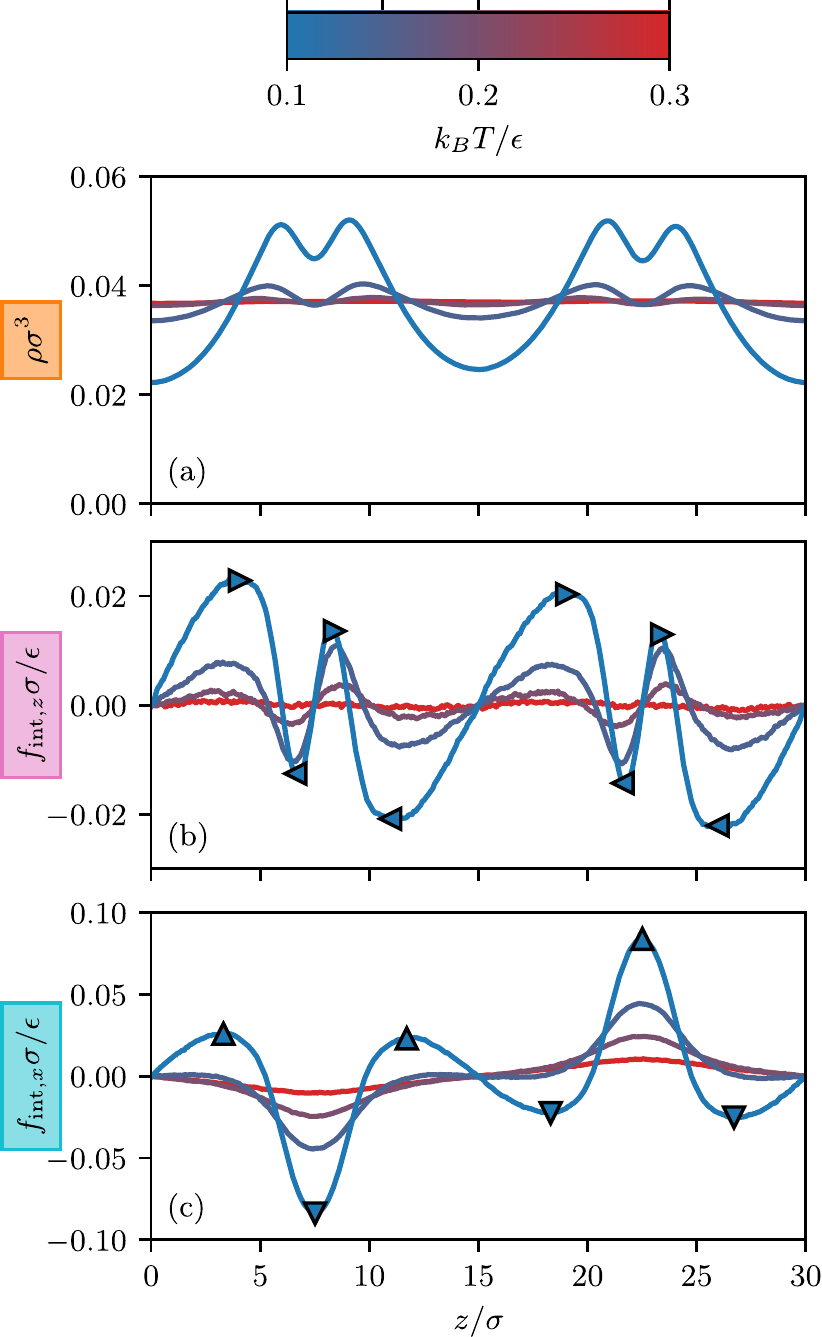}
  \caption{
    The density profile $\rho(z)$ (a) as well as the component $f_{\mathrm{int}, z}(z)$ (b) and $f_{\mathrm{int}, x}(z)$ (c) of the internal force \eqref{eq:fext} is shown.
    A constant shear amplitude of $K = 5 \epsilon / \sigma$ is maintained and the temperature is varied with values of $k_B T / \epsilon = 0.1, 0.15, 0.2, 0.3$ (indicated by ticks on the color scale).
    While $f_{\mathrm{int}, x}(z)$ acts parallel to the flow direction, $f_{\mathrm{int}, z}(z)$ constitutes a force perpendicular to the flow that leads to the observed density inhomogeneity.
    This is illustrated by arrows, which accentuate in particular the alternating direction in both the parallel and the perpendicular internal force component for low temperature.
    The onset of structural inhomogeneities in the one-body profiles is continuous and occurs rapidly for decreasing $T$ when the equilibrium percolation transition is encountered.
  }
  \label{fig:T_sweep_profiles}
\end{figure}

\begin{figure}[htb]
  \centering
  \includegraphics{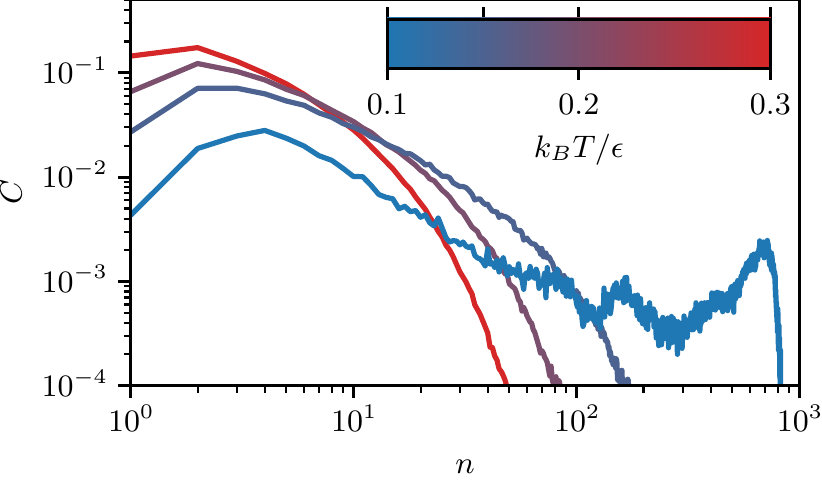}
  \caption{
    The cluster size distribution $C(n)$ is shown for different values of the temperature (indicated by ticks on the color scale) at constant shear amplitude $K = 5 \epsilon / \sigma$.
    Particles tend to form chains when driven by the shear force, cf.\ fig.~\ref{fig:system_sketch}, and the mean size of the chains grows when temperature is decreased.
    Additionally, for $k_B T = 0.1 \epsilon$, the occurence of large clusters that span across a flow channel and include up to half of the particles in the system is observed.
  }
  \label{fig:T_sweep_cluster}
\end{figure}

\begin{figure}[htb]
  \centering
  \includegraphics{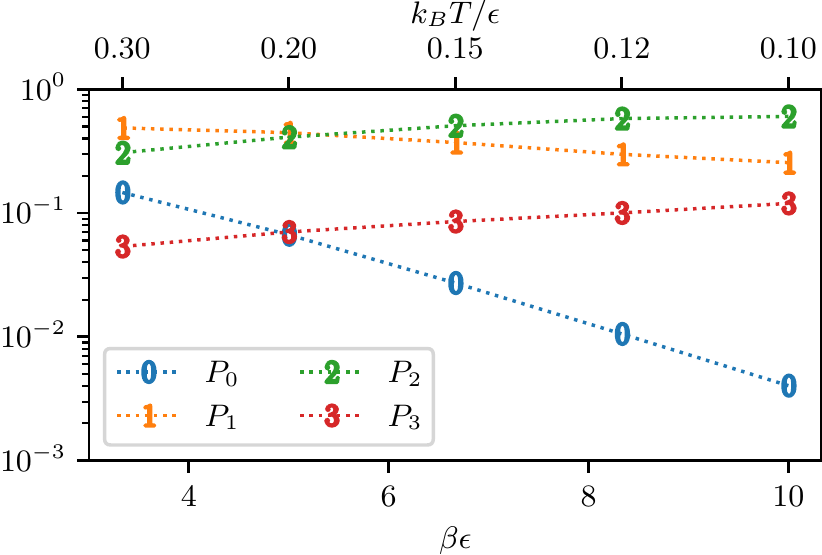}
  \caption{
    The probabilities $P_n$ of particles with a coordination number of $n = 0, 1, 2, 3$ are shown as a function of inverse temperature $\beta$ and for constant shear amplitude $K = 5 \epsilon / \sigma$.
    For low temperatures, individual particles ($n = 0$) as well as particle pairs ($n = 1$) are desorbed into the network, as both $P_0$ and $P_1$ decrease.
    The network structure is dominated by chains ($P_2$ is large) and branching is still viable, as can be deduced from the moderate value of $P_3$.
  }
  \label{fig:T_sweep_coordination}
\end{figure}

For certain state points and values of the amplitude of the external force, large variations in the one-body profiles of density $\rho(\vec{r}) = \rho(z)$ and internal force $\vec{f}_\mathrm{int}(\vec{r}) = \vec{f}_\mathrm{int}(z)$ are observed while the system retains translational symmetry in the $x$- and $y$-directions.
To investigate the onset and origin of these inhomogeneities, we first vary the temperature $T$ and maintain a large constant amplitude $K = 5 \epsilon / \sigma$ of the external force profile.
The results are shown in fig.~\ref{fig:T_sweep_profiles}.

The one-body profiles remain almost featureless for $k_B T = 0.3 \epsilon$.
At $k_B T = 0.2 \epsilon$, an inhomogeneous structure begins to appear in the internal force profiles $f_{\mathrm{int},x}(z)$ and $f_{\mathrm{int},z}(z)$.
This becomes more clearly visible as variations of the density profile $\rho(z)$ from its bulk value for $k_B T = 0.15 \epsilon$.
For $k_B T = 0.1 \epsilon$, remarkable modulations occur in all three quantities with spatial density variations of the order of the mean bulk density $\rho_b$ itself.

The emergence of structural features in the one-body profiles when decreasing temperature is rapid and continuous.
The spatial modulations are significantly stronger than those observed in (non-percolated) simple fluids \cite{Stuhlmuller2018}, and we illustrate this in Appendix \ref{appendix:LJ} via a comparison to results for the dilute Lennard-Jones fluid.
Therefore, this effect can be linked to the percolation transition in equilibrium, which sets in at similar thermodynamic state points for the considered particle model \footnote{The onset of percolation has been verified in corresponding bulk simulations. In this case, the density profile remains constant and the internal force profile vanishes within numerical accuracy}.
We support this reasoning by an investigation of the cluster size distribution $C(n)$, which gives the probability of finding a random particle in a cluster of size $n$.
As is standard, we define the agglomeration of particles into clusters to be transitive, with two particles belonging to the same cluster if their distance is below the cutoff distance $a \sigma$ of the interparticle potential.
In fig.~\ref{fig:T_sweep_cluster}, $C(n)$ is shown for varying temperature in a system sheared according to eq.~\eqref{eq:fext} with $K = 5 \epsilon / \sigma$.
One recognizes that the mean cluster size grows with decreasing temperature and that clusters span up to half of the system for $k_B T = 0.1 \epsilon$.

Additionally, to better reveal the internal structure of the clusters, we monitor the probabilities of the coordination numbers $P_n$, i.e.\ the proportion of particles having $n$ neighboring particles within the cutoff distance $a \sigma$.
The behavior of the coordination numbers $n = 0, 1, 2, 3$ is shown in fig.~\ref{fig:T_sweep_coordination} as a function of inverse temperature.
It is apparent that for low temperatures, the structure of the network is dominated by particle chains.
Branching still occurs, which interconnects the chains within the flow channels.
This shows that even in strongly sheared systems, microscopic correlations are dominated by the three-body contribution to the internal interaction potential \eqref{eq:SW}.

\subsection{Variation of external force amplitude}
\label{sec:K_sweep}

\begin{figure}[!htb]
  \centering
  \includegraphics{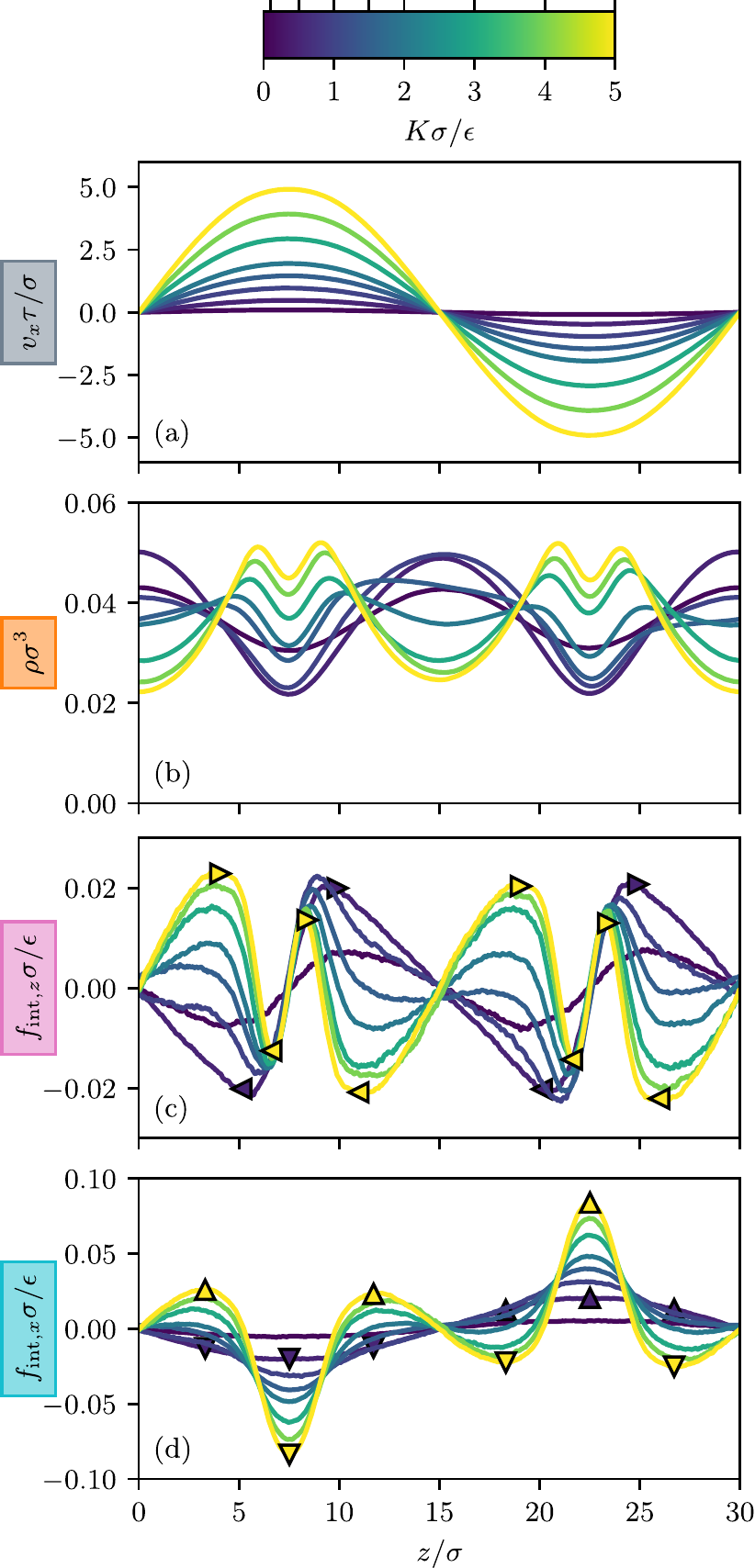}
  \caption{
    Similar to fig.~\ref{fig:T_sweep_profiles}, the one-body profiles of velocity $v_x(z)$ (a), density $\rho(z)$ (b), perpendicular internal force $f_{\mathrm{int}, z}(z)$ (c) and parallel internal force $f_{\mathrm{int}, x}(z)$ (d) are depicted.
    The temperature is now fixed to $k_B T = 0.1 \epsilon$ and the amplitude of the external force is varied with values of $K \sigma / \epsilon = 0.1, 0.5, 1, 1.5, 2, 3, 4, 5$, which are indicated by ticks on the color scale ($K = 0$ corresponds to the bulk state and is not shown).
    Arrows indicate the local direction of the internal force components for low (blue) and high (yellow) shear amplitude.
  }
  \label{fig:K_sweep_profiles}
\end{figure}

While the formation of finite-size clusters can be understood as a relic of the equilibrium percolation transition, its effect on the concrete structure of $\rho(z)$ and $\vec{f}_\mathrm{int}(z)$ turns out to be substantial and can only be explained if genuine non-equilibrium dynamics are considered.
In the following, the response of the three-body gel over a range of external force amplitudes $K$ at constant (low) temperature $k_B T = 0.1 \epsilon$ is investigated, whereby the system is driven further away from equilibrium with increasing $K$.
The corresponding one-body profiles are shown in fig.~\ref{fig:K_sweep_profiles}.

To rationalize the results, it is instructive to work on the level of forces and to consider the one-body force balance \cite{Schmidt2022}
\begin{equation}
  \label{eq:force_balance}
  \zeta \vec{v}(\vec{r}) = \vec{f}_\mathrm{int}(\vec{r}) + \vec{f}_\mathrm{ext}(\vec{r}) - k_B T \nabla \ln\rho(\vec{r}),
\end{equation}
The above relation is exact for arbitrary many-body Hamiltonians, which can be shown e.g.\ via an integrating-out of the Smoluchowski equation or in equilibrium, where $v(\vec{r}) = 0$, by an application of Noether's theorem \cite{Hermann2021}.
The external force $\vec{f}_\mathrm{ext}(\vec{r})$ is imposed in our system via eq.~\eqref{eq:fext}, and $\rho(\vec{r})$ as well as $\vec{f}_\mathrm{int}(\vec{r})$ are accessible from their microscopic definitions given in eqs.~\eqref{eq:rho_op} to \eqref{eq:fint}.
The term $-k_B T \nabla \ln\rho(\vec{r}) = \vec{f}_\mathrm{id}(\vec{r})$ on the right-hand side of eq.~\eqref{eq:force_balance} is the force arising from ideal-gas diffusion.
Further, the time dependence has been dropped as we consider a stationary state where $\vec{v}(\vec{r}) = \vec{J}(\vec{r}) / \rho(\vec{r})$ is the time-independent one-body velocity, which can be obtained from the one-body current $\vec{J}(\vec{r})$.
The current obeys the continuity equation $\partial \rho(\vec{r}, t) / \partial t = - \nabla \cdot \vec{J}(\vec{r}, t)$ and it is therefore divergence-free in the present case since the density profile is stationary.
We recall that $\vec{J}(\vec{r})$ is an average of the microscopic operator $\hat{\vec{J}}(\vec{r}) = \sum_i \vec{v}^{(i)} \delta(\vec{r} - \vec{r}^{(i)})$ and it is thus directly accessible from simulation\cite{delasHeras2019}.

We proceed similar to Ref.~\onlinecite{delasHeras2020} and distinguish between adiabatic and superadiabatic contributions to the internal force profile $\vec{f}_\mathrm{int}(\vec{r}) = \vec{f}_\mathrm{ad}(\vec{r}) + \vec{f}_\mathrm{sup}(\vec{r})$.
The adiabatic force $\vec{f}_\mathrm{ad}(\vec{r})$ is defined to be that of a reference equilibrium system, which is constructed to have the same density profile as the original non-equilibrium state.
Only the superadiabatic part $\vec{f}_\mathrm{sup}(\vec{r})$ consists of purely out-of-equilibrium forces, which hence determine both (inhomogeneous) structure and flow of a driven colloidal suspension.
By specializing to our planar geometry, we now analyze the force profiles in fig.~\ref{fig:K_sweep_profiles} to determine adiabatic and superadiabatic contributions.

Parallel to the flow, the density remains homogeneous due to translational symmetry, such that the $x$-component of its gradient vanishes.
This also implies a vanishing adiabatic force $f_{\mathrm{ad}, x}(z) = 0$.
The respective component of the internal force therefore only consists of the superadiabatic contribution, i.e.\ $f_{\mathrm{int}, x}(z) = f_{\mathrm{sup}, x}(z)$.
Hence, the $x$-component of the force balance eq.~\eqref{eq:force_balance} simplifies to
\begin{equation}
  \label{eq:force_balance_x}
  \zeta v_x(z) = f_{\mathrm{sup}, x}(z) + f_{\mathrm{ext}, x}(z),
\end{equation}
which clarifies that $f_{\mathrm{sup, x}}(z)$ plays the role of a viscous force and that it is readily available via the simulation results for $f_{\mathrm{int}, x}(z)$ shown in fig.~\ref{fig:K_sweep_profiles}.

In the $z$-direction, the density varies inhomogeneously and the internal force therefore consists of both adiabatic and superadiabatic contributions.
To distill $f_{\mathrm{sup}, z}(z)$ from the data of $f_{\mathrm{int}, z}(z)$ in fig.~\ref{fig:K_sweep_profiles} would require the construction and simulation of an appropriately chosen equilibrium system \cite{Fortini2014,delasHeras2019}, which will be considered in future work.
Nevertheless, due to the absence of driving and flow in the $z$-direction, i.e.\ $f_{\mathrm{ext}, z}(z) = 0$ and $v_z(z) = 0$, the force balance along the $z$-axis reduces to
\begin{equation}
  \label{eq:force_balance_z}
  0 = f_{\mathrm{sup}, z}(z) + f_{\mathrm{ad}, z}(z) - k_B T \frac{\partial \ln\rho(z)}{\partial z}.
\end{equation}
Therefore, the non-equilibrium force component $f_{\mathrm{sup}, z}(z)$ is necessary to stabilize the density gradient, and it can thus be referred to as a structural superadiabatic force.
Eq.~\eqref{eq:force_balance_z} also reveals that the internal force density is straightforwardly related to the derivative of the density profile due to $F_{\mathrm{int}, z}(z) = f_{\mathrm{int}, z}(z) \rho(z) = k_B T \partial \rho(z) / \partial z$, which can be utilized as a means to ``force sample'' \cite{Borgis2013,delasHeras2018a,Rotenberg2020} the density profile with a reduced variance.
Additionally, an analogous description of viscous and structural effects on the level of internal stresses is given in Appendix \ref{appendix:stress}.

In simple fluids, where the constituent particles only interact via an isotropic pair-potential, non-equilibrium viscous and structural forces have been reported to occur both in an analogous sinusoidal shear profile \cite{Stuhlmuller2018} and in more complex two-dimensional flows \cite{delasHeras2020}.
However, the emerging features of density and force profiles -- while being measurable and conceptually important -- are rather frugal especially in the quasi-one-dimensional case (cf.\ Appendix \ref{appendix:LJ} for results of the sheared Lennard-Jones fluid).
The relative variation in density is comparatively small even for moderate external force, and particles consistently accumulate in regions of low shear rate, i.e.\ at the center of the flow channels.
The superadiabatic forces possess a sinusoidal shape such that the structural force drives particles to the center of the channels.
The viscous force is Stokes-like in a broad range of shear amplitudes and it is always opposed to the flow direction.
In the following, these observations are compared to the markedly different one-body profiles of the sheared three-body gel illustrated in fig.~\ref{fig:system_sketch}.
The results of the variation of $K$ are shown in fig.~\ref{fig:K_sweep_profiles}.

For small values of the external force amplitude $K$, the density is sinusoidal in shape but the amplitude is inverted as compared to the simple fluid scenario such that particles accumulate in regions of \emph{large} velocity gradient.
When $K$ is increased, the density maxima shrink while the depletion at the center of the channels remains pronounced.
For large values of $K$, we observe that particles now tend to flee the regions of high velocity gradient.
However, the migration is not simply directed towards the center of the flow channels where the local shear rate vanishes, as would be the case in simple fluids.
Instead, the density profile develops a double-peak and retains a depletion zone right at the location of maximum flow velocity.
This behavior is reflected in the form of the internal force $f_{\mathrm{int}, z}(z)$, which progresses from a sinusoidal profile for small $K$ to a rapidly varying quantity for large $K$, thereby promoting and maintaining the observed double-peak structure of $\rho(z)$ within the flow channels.
The purely superadiabatic viscous force $f_{\mathrm{int}, x}(z)$ counteracts partially the flow for low to intermediate $K$ similar to the behavior found in simple fluids.
For large external force amplitudes, however, $f_{\mathrm{int}, x}(z)$ locally acts in the \emph{same} direction as the flow velocity at the sides of the channels, which is anomalous phenomenology for a viscous force.

The striking signal in both structural and viscous forces can be explained as a consequence of the three-body interaction \eqref{eq:SW}.
As the system is weakly sheared, particles can still percolate into a large network for the chosen temperature.
With increasing $K$, bonds are first broken in regions of maximum external force such that particles become mobile and evade these regions -- thus a density depletion zone develops.
At even larger $K$, the formation of an extensive network cannot be maintained and bonds break and dynamically rejoin across the whole system.
However, driven by the three-body term in eq.~\eqref{eq:SW}, particles still tend to develop finite-size chains, which then align parallel to the flow direction.
The mobility of the individual chains enables the migration to regions of low velocity gradient and the density profile hence inverts.
Within the flow channels, the chains organize into two lanes that are slightly offset from the center and thus lead to a double-peak structure in $\rho(z)$.
This is because their alignment parallel to the flow is driven by inhomogeneous shear rate and it can therefore only occur if $\partial v_x(z) / \partial z \neq 0$.
Yet, at the extrema of the external force profile, the gradient of the resulting flow vanishes and particle bonds are not aligned.
This explains the spatial offset of the chain formation to regions of finite velocity gradient, cf.\ fig.~\ref{fig:K_sweep_profiles}.
The arrangement of particles into aligned chains also clarifies the anomalous behavior of the viscous force $f_{\mathrm{int}, x}(z)$ that is encountered in this case and that can hence be understood as a dynamical ``drag-along''.
In summary, the inclusion of three-body terms in the interaction potential greatly affects the response of colloidal suspensions to inhomogeneous shear and results in collective effects, which influence and amplify structural and viscous forces.

\subsection{Power functional theory}
\label{sec:PFT}
We next turn to a theoretical description of the simulation results with PFT \cite{Schmidt2013,Schmidt2022} and give a brief summary of its core concepts in the following.
PFT is based on an exact variational principle that reproduces the time-dependent force balance equation
\begin{equation}
  \label{eq:force_balance_t}
  \zeta \vec{v}(\vec{r}, t) = \vec{f}_\mathrm{ad}(\vec{r}, t) + \vec{f}_\mathrm{sup}(\vec{r}, t) + \vec{f}_\mathrm{ext}(\vec{r}, t) - k_B T \nabla \ln\rho(\vec{r}, t).
\end{equation}
Thereby, the nontrivial contributions $\vec{f}_\mathrm{ad}(\vec{r}, t)$ and $\vec{f}_\mathrm{sup}(\vec{r}, t)$, which together constitute the internal force profile $\vec{f}_\mathrm{int}(\vec{r}, t)$, are made accessible via universal generating functionals of the density profile $\rho(\vec{r}, t)$ and current profile $\vec{J}(\vec{r}, t)$.
Together with the external and diffusive forces (right hand side), they are balanced by the friction of the overdamped system (left hand side).

More precisely, the adiabatic force $\vec{f}_\mathrm{ad}(\vec{r}, t)$ incorporates the functional derivative of the intrinsic excess Helmholtz free energy $F_\mathrm{exc}[\rho]$,
\begin{equation}
  \label{eq:PFT_fad}
  \vec{f}_\mathrm{ad}(\vec{r}, t) = - \nabla \frac{\delta F_\mathrm{exc}[\rho]}{\delta \rho(\vec{r}, t)},
\end{equation}
where brackets denote functional dependencies.
If one uses eq.~\eqref{eq:PFT_fad} in eq.~\eqref{eq:force_balance_t} and neglects $\vec{f}_\mathrm{sup}(\vec{r}, t)$, classical dynamical density functional theory (DDFT) \cite{Evans1979} is recovered as an uncontrolled approximation.
In the sheared three-body gel, as was shown in the previous section, the dynamics are govered by genuine out-of-equilibrium effects.
Being a purely adiabatic theory by construction, DDFT is strictly unable to reproduce or describe the observed behavior in our system \cite{DDFTPerspective}.

Instead, in order to go beyond an adiabatic description, superadiabatic forces $\vec{f}_\mathrm{sup}(\vec{r}, t)$ have to be taken into account.
Within PFT, this is made possible by functional differentiation of the superadiabatic excess power functional $P_\mathrm{exc}[\rho, \vec{J}]$,
\begin{equation}
  \label{eq:PFT_fsup}
  \vec{f}_\mathrm{sup}(\vec{r}, t) = - \frac{\delta P_\mathrm{exc}[\rho, \vec{J}]}{\delta \vec{J}(\vec{r}, t)}.
\end{equation}
The force balance eq.~\eqref{eq:force_balance_t} can then be written as
\begin{equation}
  \label{eq:PFT_force_balance}
  \begin{split}
    \zeta \vec{v}(\vec{r}, t) &= - \frac{\delta P_\mathrm{exc}[\rho, \vec{J}]}{\delta \vec{J}(\vec{r}, t)} - \nabla \frac{\delta F_\mathrm{exc}[\rho]}{\delta \rho(\vec{r}, t)} \\
                               &\qquad + \vec{f}_\mathrm{ext}(\vec{r}, t) - k_B T \nabla \ln \rho(\vec{r}, t)
  \end{split}
\end{equation}
and it involves both adiabatic and superadiabatic interparticle forces as systematically generated via the respective functionals.

Therefore, if $P_\mathrm{exc}[\rho, \vec{J}]$ and $F_\mathrm{exc}[\rho]$ are known, PFT enables the dynamical description of a system subjected to an arbitrary external force profile $\vec{f}_\mathrm{ext}(\vec{r})$ via eq.~\eqref{eq:PFT_force_balance} and the continuity equation.
This reformulation, which reduces the many-body problem to a variational principle on one-body quantities, is exact in principle.
Crucially, both $P_\mathrm{exc}[\rho, \vec{J}]$ as well as $F_\mathrm{exc}[\rho]$ are intrinsic functionals that depend only on internal interactions and further intrinsic properties of the system (e.g.\ temperature, density), but not on the externally applied force profile $\vec{f}_\mathrm{ext}(\vec{r}, t)$.
In practice, for a certain interparticle interaction potential, approximations for $P_\mathrm{exc}[\rho, \vec{J}]$ and $F_\mathrm{exc}[\rho]$ must be found, which poses a nontrivial problem.
For $P_\mathrm{exc}[\rho, \vec{J}]$, the functional dependence will in general be non-local both in space and in time (i.e.\ non-Markovian) as the history of $\rho(\vec{r}, t)$ and $\vec{J}(\vec{r}, t)$ has to be considered to obtain an accurate dynamical theory for time-dependent problems.

In the following, we use the framework of PFT to develop a model that is capable of reproducing the found anomalous behavior of the viscous force profile in the sheared three-body gel.
We focus on the viscous part because it is directly accessible in simulation (we recall that $f_{\mathrm{int}, x}(z)$ is purely superadiabatic), which hence simplifies the following considerations.
Recall also that a stationary state is considered, which implies $\nabla \cdot \vec{v}(\vec{r}) = 0$ due to the continuity equation and the chosen geometry.
Since $\rho(\vec{r})$ is time-independent, we perform a change of variables to formulate $P_\mathrm{exc}[\rho, \vec{v}]$ as a functional of the velocity profile and use $\delta / \delta \vec{J}(\vec{r}) = \rho(\vec{r})^{-1} \delta / \delta \vec{v}(\vec{r})$.
To yield an approximate explicit expression for $P_\mathrm{exc}[\rho, \vec{v}]$, a semi-local velocity gradient expansion \cite{delasHeras2018} is assumed.
Due to being in a stationary state, we can further specialize to a Markovian model such that
\begin{equation}
  \label{eq:functional_general}
  P_\mathrm{exc}[\rho, \vec{v}] = \int \diff \vec{r} \phi(\rho(\vec{r}), \nabla \vec{v}(\vec{r}))
\end{equation}
with a suitable integrand $\phi(\rho(\vec{r}), \nabla \vec{v}(\vec{r}))$.

As in Refs.~\onlinecite{Stuhlmuller2018,delasHeras2020}, we perform the general expansion up to second order in $\nabla \vec{v}(\vec{r})$ to obtain an expression for the integrand $\phi(\rho(\vec{r}), \nabla \vec{v}(\vec{r}))$.
Assuming a local dependence in space and imposing rotational invariance, this expression can be reduced to
\begin{equation}
  \label{eq:phi_visc}
  \phi(\rho(\vec{r}), \nabla \vec{v}(\vec{r})) = \frac{1}{2} \eta \rho(\vec{r})^2 (\nabla \times \vec{v}(\vec{r}))^2,
\end{equation}
where $\eta$ is the coefficient of the superadiabatic viscous response.
This coefficient depends on intrinsic properties such as interparticle potential, density and temperature, but it crucially is independent of the imposed external force profile.
Further, the value of $\eta$ only alters the magnitude of the superadiabatic force profile resulting from eq.~\eqref{eq:phi_visc}, with its shape being fully determined by the forms of $\rho(\vec{r})$ and $\vec{v}(\vec{r})$.

\begin{figure}[htb]
  \centering
  \includegraphics{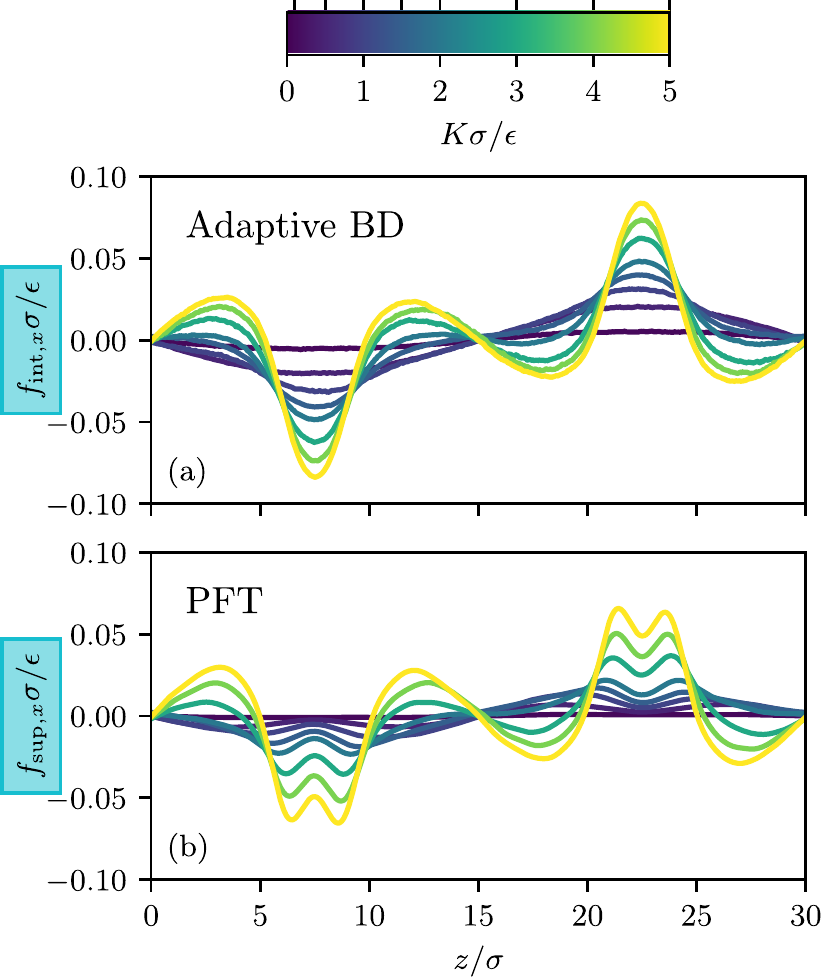}
  \caption{
    The superadiabatic viscous force $f_{\mathrm{sup}, x}(z)$ is shown as obtained from the model \eqref{eq:functional_general} and \eqref{eq:phi_visc} for $P_\mathrm{exc}[\rho, \vec{v}]$ with $\eta = 5$.
    For this, the expression \eqref{eq:fsup_for_phi_visc} is evaluated with the density profiles from the simulations in Sec.~\ref{sec:K_sweep}.
    The velocity profiles are approximated analytically by $\vec{v}(\vec{r}) \approx \vec{f}_\mathrm{ext}(\vec{r}) / \zeta$.
    When comparing the shown PFT results (b) with the adaptive BD simulation data (a) for $f_{\mathrm{int}, x}(z)$, good agreement is found.
    Solely in the center of the flow channels, the simple model for $P_\mathrm{exc}[\rho, \vec{v}]$ leads to deficiencies due to the complex behavior of $\rho(z)$ in these regions.
  }
  \label{fig:PFT}
\end{figure}

The functional minimization \eqref{eq:PFT_fsup} of eq.~\eqref{eq:functional_general} with the model integrand \eqref{eq:phi_visc} results in the superadiabatic force profile
\begin{equation}
  \label{eq:fsup_for_phi_visc}
  \begin{split}
    \vec{f}_\mathrm{sup}(\vec{r}) &= \eta \left[ \rho(\vec{r}) \nabla^2 \vec{v}(\vec{r}) - \rho(\vec{r}) \nabla (\nabla \cdot \vec{v}(\vec{r})) \right.\\
                                  &\qquad \left. - 2 (\nabla \rho(\vec{r})) \times (\nabla \times \vec{v}(\vec{r})) \right],
  \end{split}
\end{equation}
where the right hand side can be evaluated for the sheared gel.
For this, we approximate $\vec{v}(\vec{r}) \approx \vec{f}_\mathrm{ext}(\vec{r}) / \zeta$ since the magnitude of the viscous force is small compared to $\zeta \vec{v}(\vec{r})$ and take the density profile from the simulations as input.
The $x$-component of eq.~\eqref{eq:fsup_for_phi_visc} then yields the viscous force, which can be compared to the actual simulation data $f_{\mathrm{int}, x}(z)$ as shown in fig.~\ref{fig:K_sweep_profiles}.
To obtain a quantitative comparison, the value of the transport coefficient $\eta$ is fitted to match the magnitude of the simulation results universally for the considered shear amplitudes.
The superadiabatic force profiles for the viscous force within this PFT description are shown in fig.~\ref{fig:PFT}, where a value of $\eta = 5$ has been used for the viscous coefficient for all considered values of $K$.

It is apparent that $f_{\mathrm{sup}, x}(z)$ displays a double peak within the flow channels and therefore differs from the simulation results, where only a single peak is observed.
However, this inaccuracy is not surprising, since the model functional \eqref{eq:functional_general} and \eqref{eq:phi_visc} is obtained merely by an expansion in gradients of the velocity profile.
The density enters the functional only locally, and the model is thus expected to fail in regions where higher derivatives (e.g.\ the curvature) of $\rho(\vec{r})$ are significant, such as in the center of the flow channels.
To achieve better results in these regions, the integrand \eqref{eq:phi_visc} of $P_\mathrm{exc}[\rho, \vec{v}]$ could be augmented by an expansion in $\rho(\vec{r})$, which will be considered in future work.

In between the flow channels, the viscous force profiles obtained from eqs.~\eqref{eq:functional_general} and \eqref{eq:phi_visc} match the simulation results across the range of investigated shear amplitudes $K$.
Particularly, the anomalous change of sign in the viscous force, which we attribute to a dynamical drag-along of particles, is captured by the PFT model as well and it shows the same $K$-dependent behavior as in the simulation.
The successful reproduction of this phenomenon exemplifies that even simple model functionals for the excess power are capable of resolving nontrivial superadiabatic effects and that PFT is a concise framework for their systematic investigation.

\section{Conclusion and outlook}
\label{sec:conclusion_and_outlook}

In this work, we have studied the behavior of a colloidal gel modeled by the Stillinger-Weber potential \eqref{eq:SW}, where the three-body interaction \eqref{eq:u3} has been modified similar to Refs.~\onlinecite{Saw2009,Saw2011}.
The gel is subjected to a sinusoidal external shear profile.
For the numerical investigation, we have utilized adaptive BD \cite{Sammuller2021} which facilitates to carry out efficient and stable long-time simulation runs to accurately obtain the density and internal force profile in the stationary flow state.
Markedly different behavior has been encountered depending on the chosen temperature $T$ and the amplitude $K$ of the external force profile.

The simulations over a range of temperatures revealed that the effect of the equilibrium percolation transition -- which leads to the formation of an extended and dilute network under quiescent bulk conditions -- transfers to situations far from equilibrium.
Thus, while a system-spanning network is not formed for sufficiently strong inhomogeneous shear, the local arrangement of particles into finite-size chains is still viable, which we have shown via the cluster size distribution $C(n)$.
An investigation of the probabilities of the coordination numbers $P_n$ revealed that the clusters are dominated by chains, which interconnect via branching.
This clustering effect is crucial to describe the emergence of structure in the density profile $\rho(z)$ and in the parallel and perpendicular component of the internal force $\vec{f}_\mathrm{int}(z)$ with respect to the flow direction.
Note that the global temperature acts as a control parameter for the network formation in our system.
In depletion-induced gels, a similar effect could be achieved by a variation of the concentration of the depletion agent to tailor the effective attraction between colloids \cite{Cates2004,Royall2008,Royall2018}.

For an in-depth analysis, we have further split the internal force into adiabatic and superadiabatic contributions, with the latter being the driving mechanism for genuine out-of-equilibrium effects.
Due to the chosen planar geometry, the parallel component of the internal force could be associated directly with a superadiabatic viscous force.
The perpendicular component consists of both adiabatic and superadiabatic contributions instead, where the latter is needed to stabilize the emerging density inhomogeneity.

When comparing the found results of the three-body gel with known observations of colloids consisting of simpler particle types \cite{Stuhlmuller2018,delasHeras2020}, we found anomalous behavior for both viscous and structural effects.
This could be attributed to be a direct consequence of the internal three-body contributions.
The emerging density modulation is much larger in magnitude and shows a richer phenomenology than in simple fluids, as we have illustrated via a comparison to the Lennard-Jones fluid in Appendix \ref{appendix:LJ}.
In particular, the accumulation of particles can occur both in regions of high and low velocity gradient depending on the applied external force.
For large amplitudes of the latter, the formation of particle chains occurs within a double-lane near the center of the flow channels.
The superadiabatic viscous force, which generally opposes the flow direction in simple fluids, has been shown here to flip its usual counteracting direction for large $K$ in some regions of the channels.
We deduced this ``drag-along'' to be another consequence of the formation of particle chains.
Therefore, in both components of the internal force profile, collective effects are involved which substantially amplify the non-equilibrium response of the system.
As we have shown, colloidal gels are very susceptible to out-of-equilibrium phenomena, and they can hence be taken as a prototypical model for future study.

By utilizing PFT, a possible route to a coarse-grained description of the found results was given.
This was exemplified for the viscous force profile, where we have shown that a simple excess power functional suffices to reproduce the simulation results and capture the anomalous drag-along in the three-body gel.
In future work, more sophisticated model functionals will be investigated in order to alleviate some deficiencies of this simple description.
Building upon the found results, a similar analysis of the structural force profile will be considered.
This requires, however, the construction of an equilibrium reference state to perform the splitting of the respective internal force component into adiabatic and superadiabatic contributions.

In the conducted simulations, it was observed that asymmetric channel populations which persist over long time scales occur especially for intermediate values of the shear amplitude.
Hence, another objective for future work is a study of their statistics and stability, possibly being indicative of a dynamical phase transition as reported already in dense colloidal suspensions of simpler particles that exhibit flow-induced ordering or layering phenomena \cite{Brader2011,Scacchi2016}.
Further interesting research could incorporate a variation of other parameters of the Stillinger-Weber potential besides $\Theta_0$ to study their impact on the response of the driven system.
This is especially important from a practical perspective, as the tuning of microscopic interactions to yield desired material properties is a central concept of material science, which has also been applied to colloidal gels under shear \cite{Koumakis2015}.
For a quantitative prediction, hydrodynamic interactions might become relevant, and it would be useful to augment adaptive BD in this regard, possibly accompanied by efficient evaluation schemes of then correlated random increments \cite{Geyer2009,Schmidt2011}.
Additionally, going beyond the steady state and investigating time-dependent situations, such as transients in a switching protocol of the external force \cite{Treffenstadt2020}, could reveal the nature of non-equilibrium memory effects.
This is especially interesting from the view point of PFT, as memory kernels can be directly incorporated in the theory, such that time-dependent phenomena may provide further assistance in the development of accurate functionals.

\section*{Data availability statement}
The data that support the findings of this study are available from the corresponding author upon reasonable request.

\begin{acknowledgments}
  We thank Tobias Eckert and Matthias Fuchs for useful comments.
  This work was submitted as a \href{https://publishing.aip.org/publications/journals/special-topics/jcp/colloidal-gels/}{\color{blue}{special topic}} to the \href{https://aip.scitation.org/topic/special-collections/gels2022?SeriesKey=jcp}{\color{blue}{collection on colloidal gels}} in J.\ Chem.\ Phys.
  This work is supported by the German Research Foundation (DFG) via project number 436306241.
\end{acknowledgments}

\bibliography{gelation.bib}

\appendix

\begin{figure}[htb]
  \centering
  \includegraphics{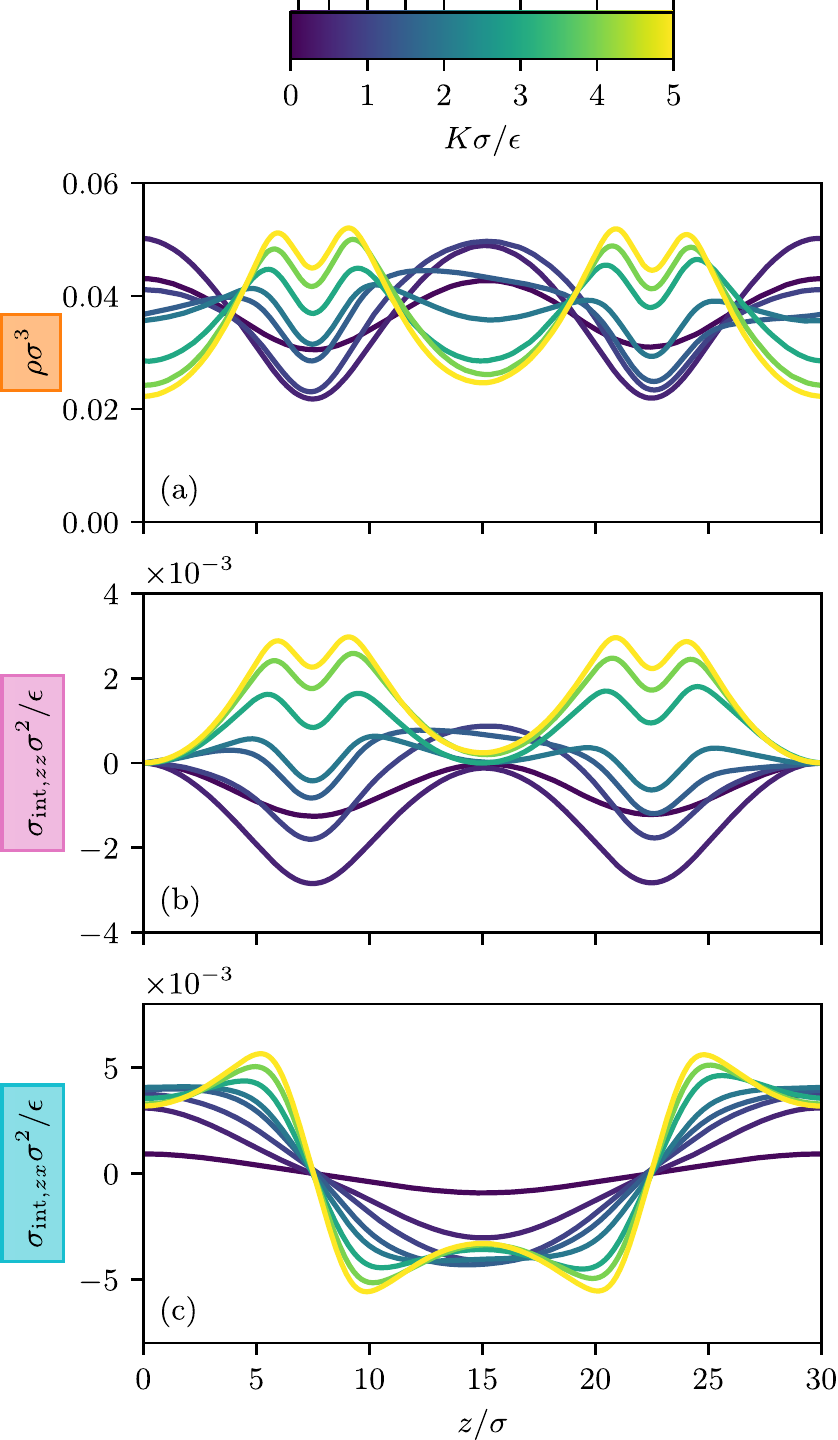}
  \caption{
    The density profile (a) as well as structural (b) and viscous (c) components of the internal stress tensor $\tensor{\sigma}_\mathrm{int}(z)$ as obtained via eqs.~\eqref{eq:stresszz} and \eqref{eq:stresszx} are shown.
    The components of the internal stress tensor are scaled by the squared particle diameter divided by the energy scale ($\sigma^2 / \epsilon$).
    A constant temperature $k_B T = 0.1 \epsilon$ is maintained and values of $K \sigma / \epsilon = 0.1, 0.5, 1, 1.5, 2, 3, 4, 5$ (indicated by ticks on the color scale) are chosen for the shear amplitude as in fig.~\ref{fig:K_sweep_profiles}.
  }
  \label{fig:K_sweep_stress}
\end{figure}

\begin{figure}[htb]
  \centering
  \includegraphics{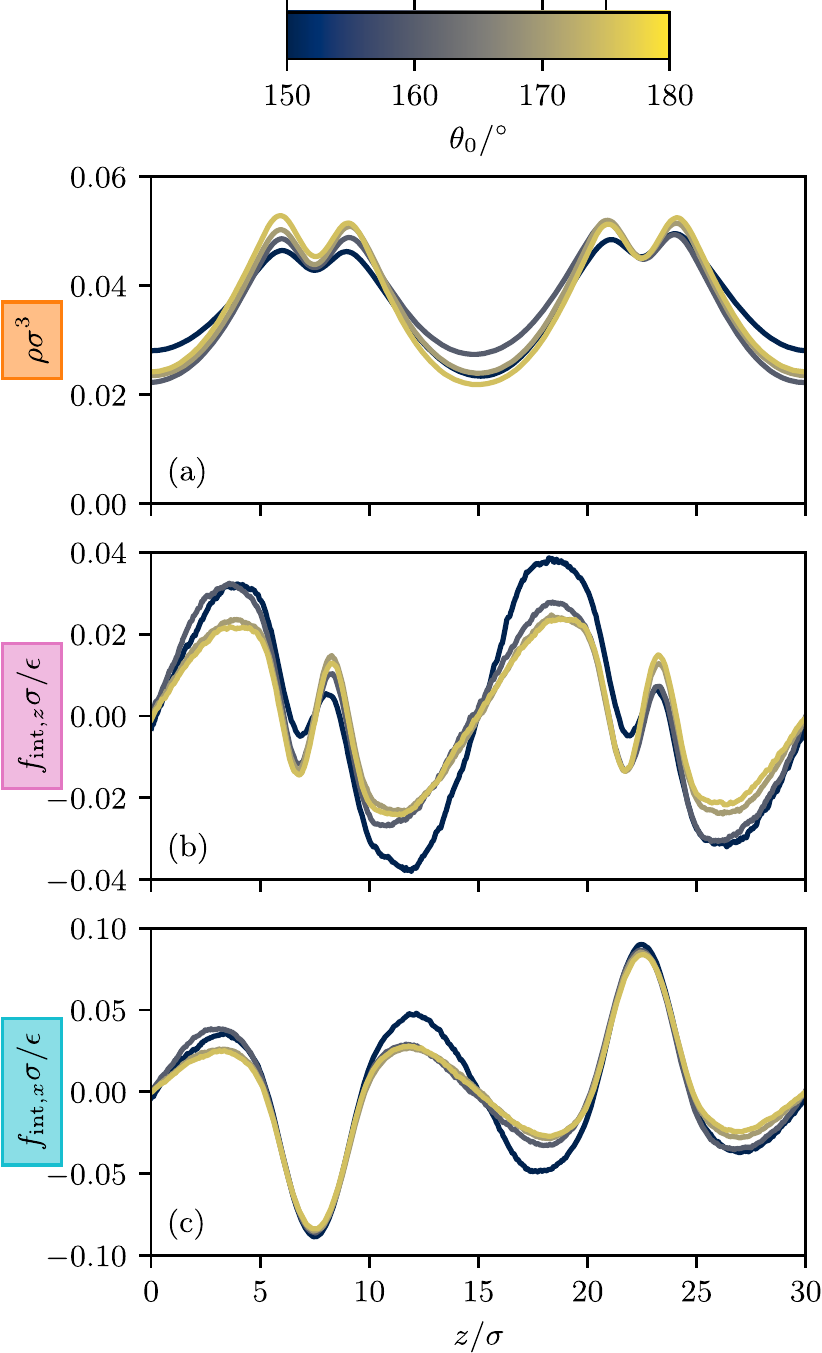}
  \caption{
    The density profile $\rho(z)$ (a) as well as the perpendicular (b) and parallel (c) component of the internal force profile $\vec{f}_{\mathrm{int}}(z)$ are shown for the sheared three-body gel with modified preferred three-body angles of $\theta_0 = 175^\circ, 170^\circ, 160^\circ, 150^\circ$ (indicated by ticks on the color scale).
    We set a temperature of $k_B T = 0.1 \epsilon$ and a shear amplitude of $K = 5 \epsilon / \sigma$.
    As network formation also occurs for the above values of $\theta_0$ and as it is the driving mechanism for the strong superadibatic response, one can observe similar behavior as for the choice of $\theta_0 = 180^\circ$ in the main text.
    Below a value of $\theta_0 = 150^\circ$, an accurate sampling of the steady state was hindered by the formation of droplets in the flow channels.
  }
  \label{fig:theta0_sweep_profiles}
\end{figure}

\begin{figure}[htb]
  \centering
  \includegraphics{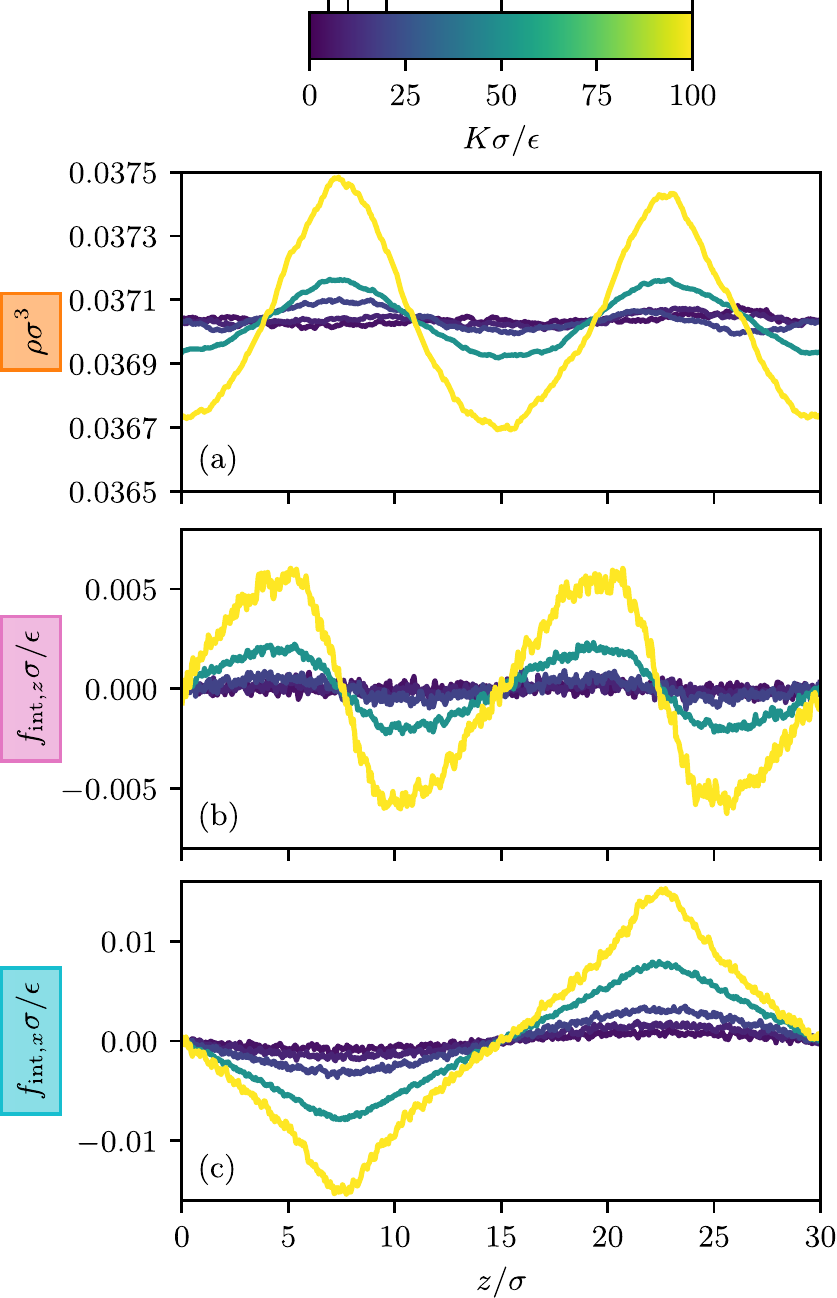}
  \caption{
    The steady state behavior of a sheared low-density Lennard-Jones fluid is shown for a temperature of $k_B T = 1.5 \epsilon$ and for various shear amplitudes $K \sigma / \epsilon = 5, 10, 20, 50, 100$ (indicated by ticks on the color scale).
    The superadiabatic response of this representative simple fluid is much weaker than in the sheared three-body gel.
    The migration of particles always occurs towards the center of the flow channels, where the velocity gradient vanishes, as can be deduced from the density profile $\rho(z)$ (a) and the perpendicular internal force profile $f_{\mathrm{int}, z}(z)$ (b).
    The viscous superadiabatic force $f_{\mathrm{int}, x}(z)$ (c) counteracts the flow direction and unlike in the three-body gel, no drag-along is observed.
  }
  \label{fig:LJ_profiles}
\end{figure}

\section{Internal stress tensor}
\label{appendix:stress}

Instead of working on the level of the force balance eq.~\eqref{eq:force_balance} directly, one can consider a similar decomposition of the stress tensor as a generator of the respective force profiles.
For this, we use the definition
\begin{equation}
  \label{eq:stress_definition}
  \nabla \cdot \tensor{\sigma}(\vec{r}, t) = \zeta \vec{J}(\vec{r}, t)
\end{equation}
of the total stress tensor $\tensor{\sigma}(\vec{r}, t)$.
To identify its internal contribution, we multiply eq.~\eqref{eq:force_balance} by the density profile $\rho(\vec{r}, t)$, which yields the force \emph{density} balance
\begin{equation}
  \label{eq:force_density_balance}
  \zeta \vec{J}(\vec{r}, t) = \vec{F}_\mathrm{int}(\vec{r}, t) + \vec{F}_\mathrm{ext}(\vec{r}, t) - k_B T \nabla \rho(\vec{r}, t).
\end{equation}
Insertion of eq.~\eqref{eq:force_density_balance} into eq.~\eqref{eq:stress_definition} and an analogous splitting then gives rise to the definition
\begin{equation}
  \label{eq:internal_stress_definition}
  \nabla \cdot \tensor{\sigma}_\mathrm{int}(\vec{r}, t) = \vec{F}_\mathrm{int}(\vec{r}, t)
\end{equation}
for the internal stress tensor $\tensor{\sigma}_\mathrm{int}(\vec{r}, t)$.

In the considered stationary state, the time dependence can be dropped.
To obtain $\tensor{\sigma}_\mathrm{int}(\vec{r})$ from the sampled force density profile $\vec{F}_\mathrm{int}(\vec{r})$ requires an integration of its spatial components according to eq.~\eqref{eq:internal_stress_definition}.
Pressure-like contributions (corresponding to integration constants) are not accessible from the force density profiles alone and require further suitable measurements in simulation \cite{Thompson2009,Shi2022}. (The standard Irving-Kirkwood \cite{Irving1950} treatment is only valid for pair-potentials.)
We omit such constants in the following and only consider relative inhomogeneities of the internal stress.
Additionally, a non-unique \cite{Schofield1982} divergence-free part of $\tensor{\sigma}_\mathrm{int}(\vec{r})$ remains undetermined from the integration of eq.~\eqref{eq:internal_stress_definition} and is set to zero.

We specialize to the planar geometry of our system, which enables a straightforward integration to obtain two relevant components of $\tensor{\sigma}_\mathrm{int}(z)$ via
\begin{align}
  \label{eq:stresszz}
  \sigma_{\mathrm{int}, zz}(z) &= \int \diff z F_{\mathrm{int}, z}(z),\\
  \label{eq:stresszx}
  \sigma_{\mathrm{int}, zx}(z) &= \int \diff z F_{\mathrm{int}, x}(z).
\end{align}

Analogous to fig.~\ref{fig:K_sweep_profiles}, where the $x$- and $z$-component of the internal force is depicted, we show results for the components $\sigma_{\mathrm{int}, zz}(z)$ and $\sigma_{\mathrm{int}, zx}(z)$ of the internal stress tensor as obtained by eqs.~\eqref{eq:stresszz} and \eqref{eq:stresszx} in fig.~\ref{fig:K_sweep_stress}.
Here, the integration constants were chosen such that $\sigma_{\mathrm{int}, zz}(z)$ vanishes at the boundaries of the box and $\sigma_{\mathrm{int}, zx}(z)$ is anti-symmetric under motion reversal ($\vec{v}(\vec{r}) \rightarrow - \vec{v}(\vec{r})$).

It is observed that $\sigma_{\mathrm{int}, zz}(z)$ reproduces the shape of the density profile, which is consistent with the considerations in the main text, cf.\ eq.~\eqref{eq:force_balance_z}.
For $\sigma_{\mathrm{int}, zx}(z)$, a sinusoidal shape is obtained at low shear amplitudes.
When increasing $K$, the $zx$-component of the internal stress tensor develops a secondary structure.
This is indicative of the non-linear response of a colloidal gel to applied shear, which manifests itself for inhomogeneous shear in an anomalous behavior of the viscous contribution.

\section{Variation of the three-body angle}
\label{appendix:theta0}

In fig.~\ref{fig:theta0_sweep_profiles}, we show illustrative results of the sheared three-body gel for different values of the preferred three-body angle $\theta_0$.
For lower values of $\theta_0$, it is increasingly difficult to obtain symmetric profiles.
We choose $\theta_0 = 150^\circ$ as the lowest value to keep away from the liquid-gas binodal and to prevent the formation of droplets within the flow channels, which hinder an accurate sampling.
It is apparent from the results that the choice of $\theta_0 = 180^\circ$ in the main text is not artificial and that similar behavior can be achieved also for lower values of $\theta_0$ as long as gelation is enforced.
When decreasing $\theta_0$, one even observes larger local forces at the sides of the flow channels (cf.\ $f_{\mathrm{int}, z}(z)$ and $f_{\mathrm{int}, x}(z)$ in fig.~\ref{fig:theta0_sweep_profiles}), as the desorption of particle strands is enhanced due to the increased ability of branching.
We refer to Refs.~\onlinecite{Saw2009,Saw2011} for an investigation of the equilibrium behavior of the three-body gel for different values of the three-body angle $\theta_0$ and the three-body interaction strength $\lambda$.

\section{Comparison to the Lennard-Jones fluid}
\label{appendix:LJ}

For comparison, we show the behavior of the truncated Lennard-Jones fluid under an analogous shear protocol as for the three-body gel.
The Lennard-Jones interaction potential only consists of the radially isotropic pairwise contribution
\begin{equation}
  \label{eq:LJ}
  u_2(r) = \begin{cases}4 \epsilon \left[ \left( \frac{\sigma}{r} \right)^{12} - \left( \frac{\sigma}{r} \right)^{6} \right], &r \leq r_c\\0, &r > r_c\end{cases}
\end{equation}
with the cutoff distance $r_c = 2.5\sigma$, and it can hence be taken as an example of a \emph{simple} fluid or colloidal suspension.

In fig.~\ref{fig:LJ_profiles}, the density profile as well as the parallel and perpendicular contribution of the internal force profile are shown for a temperature of $k_B T = 1.5 \epsilon$ and for various (large) shear amplitudes $K$.
All other system parameters are adopted from the simulations of the sheared gel, which yields the same low mean density of $\rho_b \approx 0.037 \sigma^{-3}$.
One recognizes that the superadiabatic response of the Lennard-Jones fluid differs starkly from that of the three-body gel, cf.~\ref{fig:K_sweep_profiles}.
The density inhomogeneity of the simple liquid is orders of magnitude smaller and possesses a sinusoidal shape that does not change qualitatively for different shear amplitudes.
Note that despite driving the Lennard-Jones system with much stronger external forces, the onset of notable superadiabatic effects occurs only for sufficiently large inhomogeneous shear, as opposed to the three-body gel, where a substatial density inhomogeneity develops also for low values of $K$.
In particular, no inversion of the extrema in the density profile $\rho(z)$ is observed, as was the case for the three-body gel when transitioning from low to high shear.
The internal force components reflect this situation, with both $f_{\mathrm{int}, z}(z)$ and $f_{\mathrm{int}, x}(z)$ being much smaller and showing less features than in the three-body gel.
Especially for $f_{\mathrm{int}, x}(z)$, no anomalous drag-along is observed, as the superadiabatic viscous force in the Lennard-Jones fluid always counteracts the flow.

\end{document}